\documentclass[11pt]{article}
\usepackage{amsmath,amssymb,color,epsfig,cite}
\usepackage{graphicx}
\usepackage{subfigure}
\usepackage{setspace}

\usepackage{amsfonts}

\newcommand{\hoch}[1]{$\, ^{#1}$}

\makeatletter
\@addtoreset{equation}{section}
\makeatother

\newcommand{\be}{\begin{equation}}
\newcommand{\ee}{\end{equation}}
\newcommand{\bea}{\setlength\arraycolsep{2pt} \begin{eqnarray}}
\newcommand{\eea}{\end{eqnarray}}
\newcommand{\nn}{\nonumber}

\def\ft#1#2{{\textstyle{\frac{\scriptstyle #1}{\scriptstyle #2} } }}
\def\fft#1#2{{\frac{#1}{#2}}}

\def\0{{\sst{(0)}}}
\def\1{{\sst{(1)}}}
\def\2{{\sst{(2)}}}
\def\3{{\sst{(3)}}}
\def\4{{\sst{(4)}}}
\def\5{{\sst{(5)}}}
\def\6{{\sst{(6)}}}
\def\7{{\sst{(7)}}}
\def\8{{\sst{(8)}}}
\def\9{{\sst{(9)}}}

\def\sst#1{{\scriptscriptstyle #1}}

\begin{document}



\begin{center}
{\large {\bf Exact Embeddings of JT Gravity in Strings and M-theory}}

\vspace{10pt}
Yue-Zhou Li\hoch{\dag1}, Shou-Long Li\hoch{\ddag2} and  H. L\"u\hoch{*1}

\vspace{15pt}

{\it \hoch{1}Department of Physics, Tianjin University, Tianjin 300350, China}

\vspace{10pt}

{\it \hoch{2}School of Physics, Beijing Institute of Technology, Beijing 100081, China}

\vspace{30pt}

\underline{ABSTRACT}
\end{center}

We show that two-dimensional JT gravity, the holographic dual of the IR fixed point of the SYK model, can be obtained from the consistent Kaluza-Klein reduction of a class of EMD theories in general $D$ dimensions.  For $D=4$, $5$, the EMD theories can be themselves embedded in supergravities.  These exact embeddings provide the holographic duals in the framework of strings and M-theory. We find that a class of JT gravity solutions can be lifted to become time-dependent charged extremal black holes.  They can be further lifted, for example, to describe the D1/D5-branes where the worldsheet is the Milne universe, rather than the typical Minkowski spacetime.

\vfill {\footnotesize \hoch{\dag}liyuezhou@tju.edu.cn\ \ \ \hoch{\ddag}sllee\_phys@bit.edu.cn\ \ \ \hoch{*}mrhonglu@gmail.com\ \ \ }

\pagebreak

\tableofcontents
\addtocontents{toc}{\protect\setcounter{tocdepth}{2}}


\newpage

\section{Introduction}
The AdS/CFT correspondence \cite{Maldacena:1997re} serves as a bridge that connects some conformal field theory (CFT$_d$) in $d$ dimensions and gravity in the anti-de Sitter (AdS$_D$) background in $D=d+1$ dimensions. This holographic duality was best studied between the $\mathcal{N}=4$ $D=4$ superconformal field theory and type IIB string in the AdS$_5\times S^5$ background. The duality is however expected to be applicable for wider classes of theories, possibly even beyond conformal field theories. The duality remains largely conjectural, and the simplest examples to prove this duality may be associated with the integrable models that can be solved completely. However, such models with conformal symmetries are hard to come by.

Recently, Sachdev-Ye-Kitaev (SYK) model \cite{Sachdev:1992fk,kitaevsyk,Maldacena:2016hyu}, which describes random all-to-all interactions between $N$ Majorana fermions in $0+1$ dimension, has drawn a large amount of attentions due to its integrability in the large $N$ limit. The SYK model exhibits approximate conformal symmetry in the infrared (IR) limit, suggesting that the SYK model may be a ${\rm CFT}_{1}$ at low energy. It was shown to be maximally chaotic \cite{kitaevsyk,Maldacena:2016hyu,Polchinski:2016xgd} in the sense that its out-of-time order correlators exhibit Lyapunov exponents and butterfly effects \cite{Shenker:2013pqa} and they saturate \cite{Maldacena:2015waa} the chaos bound established by black holes \cite{Shenker:2013pqa}. Therefore, the IR limit of the SYK model may have an ${\rm AdS}_2$ bulk gravity dual.

However, the naive ${\rm CFT}_1$ interpretation of the SYK model is not appropriate. For the standard CFT$_1$ interpretation, the SYK model should encode the full Virasoro algebra in the IR limit and exhibit the time reparametrization invariance.  Instead the time reparametrization symmetry is spontaneously broken into its $SL(2,\mathbb R)$ subgroup, giving rise to Goldstone zero modes \cite{Maldacena:2016hyu}. The effective theory is described by a Scwharzian action. This fact demonstrates that the model deviates from the standard ${\rm CFT}_1$ and becomes the $0+1$ dimensional ``nearly conformal field theory'' (NCFT) \cite{Maldacena:2016hyu}. Many properties of ${\rm SYK}_1$ and ${\rm NCFT}_1$ have been explored, in e.g.~ \cite{Polchinski:2016xgd,Garcia-Garcia:2016mno,Davison:2016ngz,Jian:2017unn,Stanford:2017thb,Kourkoulou:2017zaj
,Bhattacharya:2017vaz,Kitaev:2017awl,a:2018kvh,Garcia-Garcia:2018pwt,Arefeva:2018hnr,Roberts:2018mnp
,Mertens:2017mtv,Taylor:2017dly,Qi:2018rqm}. The supersymmetric generalizations \cite{Fu:2016vas,Li:2017hdt,Murugan:2017eto,Hunter-Jones:2017raw,Bulycheva:2018qcp} and higher dimensional generalizations \cite{Gu:2016oyy,Gross:2016kjj,Cai:2017vyk} were also constructed.

The NCFT$_1$ property is related to the fact that the bulk theory can not be Einstein gravity in $D=2$ dimensions. Specifically, the Einstein-Hilbert action
\be
S=\fft{1}{16\pi}\int d^2x\sqrt{-g}R
\ee
is simply a topological constant and thus gives no dynamics. Thus nontrivial $D=2$ gravity must non-minimally couple to a matter field. The simplest example is perhaps the Jackiw-Teitelboim (JT) gravity \cite{Teitelboim:1983ux,Jackiw:1984je}
\be
S=\fft{1}{16\pi}\int d^2x\sqrt{-g}\,\Phi\,(R-2\Lambda_0)\,.
\label{JT}
\ee
When $\Lambda_0$ is negative, the JT model admits the ${\rm AdS}_2$ vacuum; however, the full AdS$_2$ symmetry is broken by the nontrivial dilaton $\Phi$ \cite{Engelsoy:2016xyb}.  JT gravity may thus provide a gravity dual of the IR limit of the SYK model \cite{Maldacena:2016upp} (See appendix \ref{jttosyk} for details.), and the SYK/AdS$_2$ duality can thus be addressed in the context of JT gravity \cite{Engelsoy:2016xyb,Cadoni:2017dma,Maldacena:2016upp,Maldacena:2018lmt,Kourkoulou:2017zaj,
Jensen:2016pah,Forste:2017apw,Gonzalez:2018enk,Forste:2017kwy}.

In fact, one may consider more complicated dilaton gravities in two dimensions, which were extensively studied in the last century for addressing basic problems of quantum gravity (see, e.g. \cite{Grumiller:2002nm} for a review.) Two-dimensional gravities received new attention in the light of holography.  Almheiri and Polchinski (AP) recently introduced a general family of dilaton-gravity models
\cite{Almheiri:2014cka}
\be
S=\fft{1}{16\pi}\int d^2x\sqrt{-g}\Big(\tilde{\Phi}^2 R+\lambda(\partial\tilde{\Phi})^2-U(\tilde{\Phi})\Big)\,.
\label{AP}
\ee
For appropriate potential $U$, the theory admits AdS$_2$ vacuum with constant $\tilde \Phi^2=\Phi_0$.
One can now perform a perturbation,
\be
\tilde{\Phi}^2=\Phi_0+\Phi\,,\qquad \Phi\ll\Phi_0\,,
\ee
the effective action for the linear perturbation is then precisely the JT gravity \cite{Sarosi:2017ykf}.

The AP class of models can be obtained from higher dimensional theories such as strings and M-theory via Kaluza-Klein reductions. This provides an understanding of SYK models from the higher dimensional point of view. Indeed, many higher dimensional extremal black holes has near horizon geometries as an ${\rm AdS}_2 \times {\cal M}$ \cite{Callan:1992rs,Giddings:1992kn,Cadoni:1993rn,Cadoni:1994uf,
Michelson:1999mk,Almheiri:2011cb,Nayak:2018qej,Almheiri:2016fws}, and the near horizon region can be effectively described by $D=2$ dilaton gravities (\ref{AP}) in many situations \cite{Callan:1992rs,Giddings:1992kn,Cadoni:1993rn,Cadoni:1994uf,Sarosi:2017ykf,Kolekar:2018sba}.

However, the above embeddings of JT gravity is at the linear perturbation level. From the higher-dimensional point of view, AdS$_2$ spacetimes typically arise from the near-horizon geometry of some extremal black holes.  Thus JT gravity as a linear perturbation of the AP class describes the leading-order approximation away from the extremality. Since JT gravity itself only captures the IR behavior of SYK models, these embeddings have to deal with the subtleties associated with these two competing approximations. In this paper we seek exact embeddings of JT gravity in higher dimensions so that the leading-order approximation away from the extremality has a broader range of validity.  The simplest example is perhaps Einstein gravity with a cosmological constant in three dimensions, with the reduction ansatz being $ds_3^2 = ds_2^2 + \Phi^2 dz^2$ \cite{Taylor:2017dly,Cvetic:2016eiv,Das:2017pif,Mertens:2018fds,Gaikwad:2018dfc}.  Since AdS$_3$ emerges naturally in strings and M-theory, the reduction ansatz provides a direct link between SYK and string theories. Ref.\cite{Cvetic:2016eiv} also obtained JT gravity coupled to a Maxwell field from the STU supergravity model in four dimensions \cite{Duff:1995sm} by the Kaluza-Klein reduction on $S^2$.

In this paper, we present an alternative exact embedding of JT gravity in higher dimensions.  We construct a class of Einstein-Maxwell-Dilaton (EMD) theories in general $D$ dimensions with appropriate dilaton couplings and scalar potential. We demonstrate that JT gravity can be obtained from the EMD theories via consistent Kaluza-Klein reductions. It turns out that for $D=4$ and $D=5$, the EMD theories without the scalar potential can be embedded in supergravities, which themselves can be obtained from the Kaluza-Klein reduction of strings and M-theory.

The paper is organized as follows. In section \ref{sec:emd}, we present a class of EMD theories in $D$ dimensions and express them in the $f(R)$-frame where the manifest kinetic term of the dilaton vanishes.  We then perform consistent Kaluza-Klein reductions and show that JT gravity can indeed emerge.  In section \ref{sec:lift-sol}, we consider solutions in JT gravity and oxidize them to become solutions in the EMD theories in higher dimensions.  We find that a class of JT gravity solutions are related to the previously-known time-dependent extremal black holes. In section \ref{sec:m-theory}, we consider the EMD theories in four and five dimensions and show they are consistent truncations of the bosonic sector of supergravities and/or gauged supergravities.  This allows to embed the solutions in strings and M-theory. We conclude the paper in section \ref{sec:conclude}.  In the appendix, we give some detail review of how JT gravity can give rise to the Schwarzian action.

\section{An EMD embedding of JT gravity}
\label{sec:emd}

\subsection{A class of EMD theories}

We begin with a class of EMD theories considered in \cite{Lu:2013eoa}.  The theories consist of the metric, a scalar $\phi$, and two $U(1)$ gauge fields $A$ and ${\cal A}$.  The Lagrangian in the Einstein frame is given by
\be
{\cal L} = \sqrt{-g} \Big(R - \ft12 (\partial \phi)^2 -V(\phi)- \ft14 e^{a\phi} F^2 - \ft14 e^{b \phi} {\cal F}^2\Big)\,,\label{genemdlag}
\ee
where $F=dA$, ${\cal F}=d{\cal A}$ and the dilatonic coupling constants satisfy
\be
ab=- \fft{2(D-3)}{D-2}\,.
\ee
The scalar potential $V$ is inspired by those in gauge supergravities, given in terms of a super-potential $W$ \cite{Lu:2013eoa}
\bea
V &=& \Big(\fft{dW}{d\phi}\Big)^2 - \fft{D-1}{2(D-2)} W^2\,,\cr
W &=& \fft{\sqrt2 (D-2)^2 a}{(D-2)a^2 + 2(D-3)} g\,\Big(b e^{-\fft12 a\,\phi} - a\, e^{-\fft12 b\phi}\Big)\,.
\eea
For reasons that will become apparent, in this paper, we are particularly interested in the case with
\be
a=\sqrt{\ft{2(D-3)^2}{(D-1)(D-2)}}\,,\qquad b=-\sqrt{\ft{2(D-1)}{D-2}}\,.\label{abspecial}
\ee
The potential is thus
\bea
V &=&-(D-1) g^2 \Big( e^{-\fft{2(D-3) \phi}{\sqrt{2(D-1)(D-2)}}} + (D-3) e^{\fft{2\phi}{\sqrt{2(D-1)(D-2)}}}
\Big)\cr
&=&-(D-1) g^2 \Big( e^{- a\phi} +(D-3) e^{-\fft12 (a + b)\phi}\Big)\,.
\eea
Note that if we set $A=0$ and also turn off the scalar potential, the remainder of the theory is simply the Kaluza-Klein theory with $\cal A$ being the Kaluza-Klein vector. The EMD theories were inspired by gauged supergravities.  In $D=4$ and 5, the Lagrangians are the consistent truncations of the bosonic sector of the respective gauged STU models. Their embeddings in M-theory and type IIB strings via Kaluza-Klein sphere reductions were given in \cite{Cvetic:1999xp}.

We now make a constant shift of $\phi$, and redefine
\bea
&&\phi=\tilde \phi + c\,,\qquad A_\mu = \tilde A_\mu e^{-\fft12 a c}\,,\qquad
{\cal A}_\mu = \tilde {\cal A}_\mu e^{-\fft12 b c}\,,\nn\\
&&g^2 e^{-a c} = g_2^2\,,\qquad e^{-\fft12 (a + b) c} g^2= g_1^2\,.
\eea
Dropping the tilde, the Lagrangian takes the same form as (\ref{genemdlag}), but with $V$ now given by
\be
V=-(D-1) \Big( g_2^2 e^{- a\phi} +(D-3) g_1^2 e^{-\fft12 (a + b)\phi}\Big)\,.\label{g1g2spot}
\ee
This allows us to set the parameter $g_1$ and $g_2$ to zero independently.

\subsection{Conformal transformation}

We now make a conformal transformation \cite{Liu:2012ed}
\be
g_{\mu\nu} = e^{-\sqrt{\fft2{(D-1)(D-2)}}\,\phi}\, \tilde g_{\mu\nu}\,,\label{conftrans}
\ee
the Lagrangian, after dropping the tildes, becomes
\be
{\cal L} = \sqrt{-g} \Big(\Phi \big(R+ (D-1)(D-3)g_1^2\big)- \ft14\Phi^{-1} F^2 +
\Phi^3\big((D-1)g_2^2 - \ft14 {\cal F}^2\big)\Big )\,.\label{frframe}
\ee
where
\be
\Phi=e^{-\sqrt{\fft{D-2}{2(D-1)}}\,\phi}\,.
\ee
The conformal transformation (\ref{conftrans}) is such that the dilaton's kinetic term is absent and the equation of motion for $\Phi$ becomes algebraic.  This can be generally done in supegravities or gauged supergravities and such a conformally transformed theory was referred as the $f(R)$-version of supergravity in \cite{Liu:2012ed}.  In this paper we shall also refer (\ref{frframe}) as gravity in the $f(R)$-frame.

In order to make contact with JT gravity through dimensional reduction, we take ${\cal A}=0$ and $g_2=0$, the resulting EMD theory in the $f(R)$-frame is
\be
{\cal L} = \sqrt{-g} \Big(\Phi \big(R+ (D-1)(D-3)g_1^2\big)- \ft14\Phi^{-1} F^2\Big )\,.\label{theemd}
\ee
In four and five dimensions, the theories can be obtained from taking appropriate limit of the bosonic sector of the STU gauged supergravity models.  When $g_1=0$, they can be truncated consistently from supergravities and hence can be embedded in string and M-theory.

\subsection{JT gravity from Kaluza-Klein reduction}

In this subsection, we show that JT gravity can be obtained from this EMD theory (\ref{theemd}) by the consistent Kaluza-Klein reduction.  The internal space is taken to be $(D-2)$-dimensional Einstein-space $d\Omega_{D-2,k}^2$ with $R_{ij}=(D-3) k\, g_{ij}$, where $k=-1,0,1$.  The corresponding Ricci scalar is given by
\be
R_k =(D-2)(D-3)k\,.
\ee
In order for the reduction to be consistent, we take all the singlet of the isometry group of the internal space.  The reduction ansatz is thus given by
\be
ds_D^2=ds_2^2+\varphi(x)^2 d\Omega_{D-2,k}^2\,,\qquad A_\mu=A_\mu(x)\,,\qquad \Phi=\Phi(x)\,,
\ee
where $\mu=0,1$ and $x^\mu$ are respectively the indices and coordinates of the metric $ds_2^2$. The Kaluza-Klein reduction in the Einstein frame down to dimensions higher than or equal to three was obtained in
\cite{Bremer:1998zp}. We find that the reduced Lagrangian from (\ref{theemd}) in two dimensions is
\bea
\mathcal{L}_2&=&\sqrt{-g}\,\Phi\varphi^{D-2}\Big(R+\varphi^{-2}R_k-
(D-2)(D-3)\varphi^{-2}(\partial\varphi)^2
-2(D-2)\varphi^{-1}\Box\varphi
\cr &&+(D-1)(D-3)g_1^2-\ft{1}{4}\Phi^{-2} F^2\Big)\,.
\eea
The equation of motion associated with $A$ is
\be
\nabla_{\mu}\big(\Phi^{-1}\varphi^{D-2}{F}^{\mu\nu}\big)=0\,,
\ee
which can be solved by
\be
F=\lambda\Phi\varphi^{-(D-2)}\epsilon_{(2)}\,,
\label{elec}
\ee
where $\lambda$ is an integration constant associated with the electric charge and $\epsilon_{(2)}$ is the volume 2-form of $ds_2^2$. The variations associated with $\Phi$ and $\varphi$ yield
\bea
0&=& R+\varphi^{-2}R_k-(D-2)(D-3)\varphi^{-2}(\partial\varphi)^2
-2(D-2)\varphi^{-1}\Box\varphi\cr
&&
+(D-1)(D-3)g_1^2
+\ft{1}{4}\Phi^{-2}{F}^2\,,\label{eomscalars1}\\
0&=& 4R_k \Phi^2+(D-2)\Big(\varphi^2 {F}^2+4\Phi(-\Phi\varphi\Box\varphi+\varphi^2\Box\Phi+
(D-3)(\varphi\nabla_{\mu}\varphi\nabla^{\mu}\Phi \cr &&-\Phi(\partial\varphi)^2))\Big)\,.\label{eomscalars2}
\eea
Finally, the equation of motion associated with the variation of $g_{\mu\nu}$, when (\ref{eomscalars1}) and (\ref{eomscalars2}) are applied, is given by
\bea
&&\Phi^2\varphi R_{\mu\nu}+\Phi\big(\varphi\Box\Phi+(D-2)\nabla_{\rho}\varphi\nabla^{\rho}\Phi\big)g_{\mu\nu}+
\ft{1}{4}\varphi({F}^2 g_{\mu\nu}-2{F}_{\mu\rho}{F}_{\nu}^{\rho})
\cr &&-\Phi\big(\varphi\nabla_{\mu}\nabla_{\nu}\Phi+
(D-2)\Phi\nabla_{\mu}\nabla_{\nu}\varphi\big)=0\,.
\eea
Taking the trace yields
\be
\Phi \varphi R+\varphi\Box{\Phi}+2(D-2)\nabla_{\rho}\Phi\nabla^{\rho}\varphi-(D-2)\Phi\Box\varphi=0\,.\label{trace}
\ee
This equation, together with the two scalar equations (\ref{eomscalars1}) and (\ref{eomscalars2}), imply that we can take $\varphi=\varphi_0$ to be constant, provided that the charge parameter $\lambda$ is
\be
\lambda^2 = (D-1)(D-3) (k + g_1^2 \varphi_0^2) \varphi_0^{2(D-3)}\,.
\label{conslam}
\ee
The remainder equations can be summarized as
\be
R-2\Lambda_{0}=0\,,\qquad R_{\mu\nu}\Phi+g_{\mu\nu}\Box\Phi-\nabla_{\mu}\nabla_{\nu}\Phi=0\,.
\label{eomconsvp}
\ee
where
\be
\Lambda_{{0}}=-\ft14(D-3)\big(k(D-3)\varphi_0^{-2}
+(D-1)g_1^2\big)\,.\label{Lambda0gen}
\ee
It is now straightforward to see that the equations (\ref{eomconsvp}) can be derived from the action of JT gravity (\ref{JT}).  In appendix \ref{jttosyk}, we review how the JT action, together with the bounary terms of the Gibbons-Hawking type and the holographic counterterm, give rise to the Schwarzian action in the appropriate AdS$_2$ background. (A more general argument for how the Schwarzian action arises from AP models was presented in \cite{Cvetic:2016eiv}.)

To conclude, JT gravity with (\ref{Lambda0gen}) can be obtained from the Kaluza-Klein reduction of the $D$-dimensional EMD theory (\ref{theemd}) and the consistent reduction ansatz is
\be
ds_D^2 = ds_2^2 + \varphi_0^2 d\Omega_{D-2,k}^2\,,\qquad
F=\sqrt{(D-1)(D-3) (k + g_1^2 \varphi_0^2)}\,\varphi_0^{-1}\,\Phi\,\epsilon_{(2)}\,.
\ee
In the case of $D=3$, nontrivial results require the absorbing of the $(D-3)$ factor into $g_1^2$.  It is instructive simply to introduce $\tilde g_1^2 = (D-3) g_1^2$ and declare that $\tilde g_1^2$ is non-vanishing in $D=3$.  To be specific, we see that the $D=3$ theory in the Einstein frame is
\be
{\cal L}_3 = \sqrt{-g} \big(R - \ft12 (\partial\varphi)^2 + 2 \tilde g_1^2 e^{\varphi} - \ft14 F^2\big)\,.
\ee
In the $f(R)$-frame, it becomes
\be
{\cal L}_3 = \sqrt{-g} \Big[\Phi(R + 2 \tilde g_1^2) - \ft14 \Phi^{-1} F^2\Big]\,.
\ee
The reduction ansatz from $D=3$ to $D=2$ is
\be
ds_3^2 = ds_2^2 + \varphi_0^2\,dz^2\,,\qquad F=\sqrt2\tilde g_1 \Phi\,\epsilon_\2\,.
\ee
The resulting two-dimensional theory is then the JT theory (\ref{JT}) with $\Lambda_0=-\ft12 \tilde g_1^2$. Note that equations of motion of the resulting JT gravity are independent of the constant parameter $\varphi_0$.

The $D=3$ case perfectly illustrated the difference between our embedding of JT gravity in higher dimensions and those discussed previously in literature. In \cite{Taylor:2017dly}, (also see \cite{Cvetic:2016eiv,Das:2017pif,Mertens:2018fds,Gaikwad:2018dfc},) the higher-dimensional theory is pure AdS gravity and the scalar in JT gravity arises as the radius of the compactifying circle $z$. The running of the JT scalar is driven by this breathing mode of the internal space.  On the other hand, in our embedding, the internal radius is fixed consistently by the equations of motion to be a constant.  The JT scalar is a direct descendant of the dilaton in higher dimensions.  Both embeddings are possible due to the fact that JT gravity is not conformal in that it has a running dilaton, which may arise directly from the higher-dimensional theory or from the modulus parameter of the compactifying space in a theory that has no scalar.

\section{Oxidations and time-dependent black holes}
\label{sec:lift-sol}

In the previous section, we demonstrate that JT gravity in two dimensions can be obtained from the consistent Kaluza-Klein reduction of the EMD theory (\ref{theemd}) on an internal Einstein space.  This allows us to oxidize all the two-dimensional solutions to higher dimensions. In particular, we find that some special two-dimensional solutions become the decoupling limit of time-dependent extremal black holes.

\subsection{Oxidation of the solutions}

The two-dimensional metric is Einstein with a negative cosmological constant.  We can thus take the metric to be
\be
ds^2=-f(r)dt^2+\fft{1}{f(r)}dr^2
\ee
Then equations (\ref{eomconsvp}) of JT gravity admit the locally AdS$_2$ solution
\bea
f=-\Lambda_0 r^2 \,,\qquad \Phi=\fft{\alpha}{r}+(\beta+\gamma t-\Lambda_0^2\alpha t^2)r
\label{solution}
\eea
where $\alpha,\beta,\gamma$ are integral constants. The theory also admits the black hole solution
\bea
ds^2 &=& \fft{1}{z^2}\Big(\fft{dz^2}{f} - f\,dt^2\Big)\,,\qquad f=1-\fft{z^2}{z_0^2}\cr
\Phi &=& -\fft{2c_2 z_0^2}{z} (1-\sqrt{f}) + \fft{\sqrt{f}}z\Big(z_0 \big(c_1 \sinh (\ft{t}{z_0})-2 c_2 z_0\big)+\big(2 c_2 z_0^2+c_0\big) \cosh(\ft{t}{z_0})
\Big)\,.
\label{solution2}
\eea
For $g_1=0$, corresponding $V=0$ in $D$ dimensions, it follows from (\ref{Lambda0gen}) that we must require $k=1$, and hence
\be
\Lambda_0=-\ft14(D-3)^2\varphi_0^{-2}\,.
\ee
In this case, the two-dimensional solutions can be lifted to become $D$-dimensional ones.  In the Einstein frame, they take the form
\bea
&& d\hat{s}^2=\Phi^{\ft{2}{D-2}}\Big(-f\, dt^2+\fft{1}{f}dr^2+\varphi_0^2 d\Omega_{D-2}^2\Big)\,,\cr
&&  F=\sqrt{(D-1)(D-3)}\,\varphi_0^{-1}\Phi\, dt\wedge dr\,,\qquad \phi=-\sqrt{\ft{2(D-1)}{D-2}}\log\Phi\,.
\label{emdlift}
\eea
For $g_1\ne 0$, we can have all $k=1,0,-1$, and $\Lambda_0$ is given by (\ref{Lambda0gen}). The $D$-dimensional solutions become
\bea
&& d\hat{s}^2=\Phi^{\ft{2}{D-2}}\Big(-f\, dt^2+\fft{1}{f}dr^2+\varphi_0^2 d\Omega_{D-2,k}^2\Big)\,,\cr
&&  F=\sqrt{(D-1)(D-3)(k + g_1^2 \varphi_0^2)}\,\varphi_0^{-1}\Phi\, dt\wedge dr\,,\qquad \phi=-\sqrt{\ft{2(D-1)}{D-2}}\log\Phi\,.
\label{emdlift1}
\eea

\subsection{Time-dependent extremal black holes}

The general EMD theory (\ref{genemdlag}) admits a class of charged black hole solutions \cite{Lu:2013eoa}.  For the dilaton coupling choice (\ref{abspecial}) and vanishing scalar potential, the charged extremal black hole is given by
\bea
ds^2 &=& - H^{-\fft{D-1}{D-2}} {\cal H}^{-\fft{D-3}{D-2}} dt^2 + H^{\fft{D-1}{(D-2)(D-3)}}
{\cal H}^{\fft{1}{D-2}} \Big(dr^2 + r^2 d\Omega_{D-2}^2\Big)\,,\nn\\
A&=& \sqrt{\ft{D-1}{D-3}}\, H^{-1} dt\,,\qquad {\cal A} = {\cal H}^{-1} dt\,,\qquad
\phi = \ft12 \ft{D-1}{D-3} a \log H + \ft12 b \log {\cal H}\,,\label{bhsol}
\eea
where $H= 1 + q/r^{D-3}$ and ${\cal H}=1 + \tilde q/r^{D-3}$ are harmonic functions in the transverse Euclidean space.  It was observed in \cite{Gibbons:2005rt} that the harmonic function ${\cal H}$ can have a linear time-dependence, namely
\be
{\cal H}=h t + \fft{\tilde q}{r^{D-3}}\,.\label{calH}
\ee
Note that here only ${\cal H}$, not $H$ can allow such a linear time dependence.
After the conformal transformation (\ref{conftrans}), the solution becomes
\bea
ds^2 &=& -\fft{dt^2}{H {\cal H}} + H^{\fft2{D-3}} \Big(dr^2 + r^2 d\Omega_{D-2}^2\Big)\,,\cr
A&=& \sqrt{\ft{D-1}{D-3}}\, H^{-1} dt\,,\qquad {\cal A} = {\cal H}^{-1} dt\,,\qquad
\Phi=\sqrt{\fft{\cal H}{H}}\,.
\eea

To make contact with the solutions in JT gravity, we set $\tilde q=0$ and furthermore we take the decoupling limit with 1 in $H$ dropped.  The solution becomes
\bea
ds^2 &=&  -\fft{r^{D-3}}{h r_0^{D-3}} d\tilde t^2 +\fft{r_0^2}{r^2} dr^2 + r_0^2
d\Omega_{D-2}^2\,,\nn\\
\Phi &=& \ft12 \tilde t \sqrt{\ft{h r^{D-3}}{r_0^{D-3}}}\,,\qquad q=r_0^{D-3}\,,
\eea
where we have redefined the time coordinate by $\tilde{t}=2\sqrt{t}$. Comparing with the Kaluza-Klein reduction ansatz (\ref{emdlift1}), we see that $\varphi_0=r_0$. Performing the Kaluza-Klein reduction on the $S^{D-2}$ sphere, and redefining the $r$ coordinate and parameters by
\be
\tilde{r}=\fft{r^{\fft{D-3}{2}}}{\sqrt{\gamma(D-3)r_{0}^{D-4}}}\,,\qquad h=\fft{4\gamma r_0}{D-3}\,,
\ee
we arrive, after dropping all the tildes, at the two-dimensional solution (\ref{solution}) of JT gravity with $\alpha=0$.

\section{Embeddings in strings and M-theory}
\label{sec:m-theory}

In the previous sections, we show that JT gravity can be obtained from consistent Kaluza-Klein reduction of a class of EMD theory (\ref{theemd}) in general $D$ dimensions. In four and five dimensions, the EMD theories can be embedded in supergravities, allowing exact embeddings of JT gravity in strings and M-theory. This will provide a better understanding of the time-dependence of the solutions.

\subsection{The $D=5$ theory}

We first consider the $D=5$ EMD theory (\ref{theemd}).  It follows from the discussion in section \ref{sec:emd} that in the Einstein frame, the Lagrangian is given by
\be
{\cal L}_5 = \sqrt{-g} \big(R - \ft12(\partial\phi)^2 - \ft14 e^{\fft{2}{\sqrt6}\phi} F^2 + 8 g_1^2 e^{-\fft1{\sqrt6}\phi}\big)\,,\label{d5emd}
\ee
For $g_1=0$, the theory can be embedded into the $U(1)^3$ supergravity with two field strengths set equal whilst the third set zero.  The theory can be obtained from ${\cal N}=(1,0)$ supergravity in $D=6$ via Kaluza-Klein reduction on $S^1$ \cite{Lu:1995cs}.  The relevant six-dimensional theory is Einstein theory coupled to a self-dual 3-form $F_\3={*F}_\3$, with
\be
R_{MN} = \ft14 F_{MPQ}F_{N}{}^{PQ}\,.
\ee
The (truncated) reduction ansatz is given by
\be
ds_6^2 = e^{\fft1{\sqrt6}\phi} ds_5^2 + e^{-\fft{3}{\sqrt6}\phi} dz^2\,,
\qquad F_\3=\ft{1}{\sqrt2}\big(F\wedge dz + {*_5F}\big)\,.\label{d6tod5}
\ee
The $D=5$ time-dependent extremal black hole (\ref{bhsol}) with $\tilde q=0$ becomes
\bea
ds_6^2 &=& H^{-1} \Big( -{\cal H}^{-1} dt^2 + {\cal H} dz^2\Big) + H (dr^2 + r^2 d\Omega_3^2)\,,\cr
F_\3 &=& -\fft{H'}{H^2} dt\wedge dz \wedge dr + 2q\Omega_\3\,,\qquad H=1 + \fft{q}{r^2}\,,\qquad {\cal H}=h t\,.
\eea
This is the self-dual string, with the flat worldsheet metric being the two-dimensional Milne universe, namely
\be
-{\cal H}^{-1} dt^2 + {\cal H} dz^2 = -h^{-1} d\tilde t^2 + \ft14 h \tilde t^2 dz^2\,,
\ee
where $t=\fft14 \tilde t^2$.  The solution can be further lifted to become the intersecting D1/D5 system, or M2/M5-branes. The near horizon geometry has an AdS$_3$ factor whose boundary is the two-dimensional Milne universe rather than the Minkowski spacetime.

    We can also lift the full two-dimensional solution (\ref{solution}) back to $D=6$ directly, and we
find
\bea
ds_6^2 &=& -\varphi_0^2 r^2 dt^2 + \Phi(r,t)^2 dz^2 + \fft{dr^2}{\varphi_0^2 r^2} + \varphi_0^2 d\Omega_3^2\,,\cr
F_\3 &=& 2\varphi_0^{-1} \Phi\, dt\wedge dr\wedge dz + 2\varphi_0^2\, \Omega_\3\,.
\eea
In particular, when $\alpha=0$, this is precisely the near-horizon geometry of the Milne self-dual string discussed above.

\subsection{The $D=4$ theory}

The EMD theory (\ref{theemd}) in four dimensions in the Einstein frame is given by
\be
{\cal L}_4=\sqrt{-g} \big(R - \ft12(\partial\phi)^2 - \ft14 e^{\fft{1}{\sqrt3}\phi} F^2 + 3 g_1^2 e^{-\fft1{\sqrt3}\phi}\big)\,,\label{d4emd}
\ee
For $g_1=0$, the theory can be embedded into the STU supergravity model, with three equal gauge potentials set equal and the fourth one set zero. The embedding of JT gravity in this theory is part of more general Kaluza-Klein reductions obtained in \cite{Cvetic:2016eiv}. The $D=4$ theory $(g_1=0)$ can be obtained from the $S^1$ reduction of minimal supergravity in five dimensions, with
\be
ds_5^2 = e^{-\fft{1}{\sqrt3}\phi} ds_4^2 + e^{\fft2{\sqrt3} \phi} dz^2\,.\label{d5tod4}
\ee
The five-dimensional Maxwell field descends down to four dimensions directly.  The time-dependent black hole solution becomes
\bea
ds_5^2 &=& - H^{-2} dt^2 + H \Big( h t(dr^2 + r^2 d\Omega_2^2) + \fft{1}{ht} dz^2\Big)\,,\cr
A &=& \sqrt3 H^{-1} dt\,.\label{d5timesol}
\eea
Turning off the charge by setting $q=0$, and hence $H=1$, the metric describes a Kasner-type cosmological solution.  We can also lift the solution (\ref{emdlift}) to five dimensions and find
\bea
&& ds_{5}^2=\Phi^2\Big(-\fft{r^2}{4\varphi_0^2}dt^2+\fft{4\varphi_0^2}{r^2}dr^2+\varphi_0^2 d\Omega_{2}^2\Big)+\Phi^{-2}dz^2
\cr && F=\sqrt{3}\,\varphi_0^{-1}\Phi(r,t)\,dt\wedge dr\,,
\eea
where $\Phi$ is given in (\ref{solution}).  This solution with $\alpha=0$ is precisely the decoupling limit of (\ref{d5timesol}) with the 1 in $H$ dropped.

The $D=4$ EMD theory with $g_1=0$ can also be embedded in M-theory with the reduction ansatz given by
\bea
ds_{11}^2 &=& e^{-\fft{1}{\sqrt3}\phi} ds_4^2 + e^{\fft2{\sqrt3} \phi} dz^2 + d\Sigma_6^2\,,\cr
A_\3 &=& \ft{1}{\sqrt3} A\, dt\wedge I_\2\,,
\eea
where $d\Sigma_6^2$ is Euclidean or Ricci-flat Calabi-Yau space with a harmonic 2-form $I_\2$.  The oxidized solution can be viewed as the M2/M2/M2 intersection, or M2-brane wrapping around the Calabi-Yau 2-cycles.

We have examined the EMD theories (\ref{theemd}) for $D=4$ and 5, with $g_1=0$.  These theories can be embedded into supergravities.  This provides many routes of JT gravity to strings and M-theory, since there are many different ways of embedding of these EMD theories in the fundamental theories. (See e.g.~\cite{Lu:1995yn}.) The higher dimensional solutions are related to M-branes, D-branes and their intersections, see e.g.~\cite{Tseytlin:1996bh,Behrndt:1996pm,Lu:1997hb}

\subsection{$D=4,5$ theories with $g_1\ne0$}

When $g_1\ne 0$, the theory can also be obtained from taking an appropriate limit of gauged supergravities, by taking $g_2=0$ in the scalar potential (\ref{g1g2spot}).  However, the sphere reduction ans\"atze constructed for $g_1g_2\ne0$ \cite{Cvetic:1999xp} do not appear to allow the $g_2=0$ limit.  It turns out that although the $D=4,5$ theories with $g_2=0$ and $g_1\ne 0$ cannot be obtained M-theory or strings directly from sphere reductions, they can be obtained from $S^1$ reduction of gauged supergravities that can be reduced from higher dimensions on spheres.

   For example, the Lagrangian for the bosonic sector of minimal gauged supergravity in five dimensions
is
\be
{\cal L}_5 =\sqrt{-g} (R - \ft14 F^2 + 12 g^2) + \ft{1}{12\sqrt3} \varepsilon^{\mu\nu\rho\sigma\lambda}
F_{\mu\nu} F_{\rho\sigma} A_\lambda\,.\label{d5gauged}
\ee
For the pure electric ansatz we consider in this paper, the last $FFA$ term can be dropped.  The reduction ansatz (\ref{d5tod4}) with the $F$ descending directly gives rise to precisely the $D=4$ EMD theory (\ref{d4emd}), with $g_1=2g$.  The lifting of the (\ref{solution}) leads to
\bea
&& ds_5^2=-\fft{1}{4}(k\varphi_0^2+3g_1^2)r^2dt^2+\fft{4}{(k\varphi_0^2+3g_1^2)r^2}dr^2+\varphi_0^2 d\Omega_{2,k}^2+\Phi^2 dz^2
\cr && F=\sqrt{3(k+g_1^2\varphi_0^2)}\varphi_0\,\Omega_{(2),k}
\eea
The five-dimensional gauged supergravity (\ref{d5gauged}) can itself be obtained from type IIB supergravity on $S^5$, and the reduction ansatz can be found in \cite{Cvetic:1999xp}.  Its (singular) embedding in M-theory was also obtained \cite{Colgain:2014pha}.  The effective cosmological constant $\Lambda_0$ (\ref{Lambda0gen}) in JT gravity now has contributions from D3-brane charges as well as the gauged supergravity R-charges associated with rotations of D3-branes.

The $D=5$ theory can also be obtained from $D=6$ Einstein gravity coupled to cosmological constant and a self-dual 3-form
\be
R_{MN} = \ft14 F_{MPQ}F_{N}{}^{PQ}-5 g^2\, g_{MN}\,,\qquad F_\3={*F}_\3\,.\label{d6selfdual}
\ee
The reduction ansatz (\ref{d6tod5}) gives precisely (\ref{d5emd}) with $g_1^2=5g^2/2$. The resulting solution is given by
\bea
&& ds_6^2=-(k\varphi_0^{-2}+2g_1^2)r^2 dt^2+\fft{1}{(k\varphi_0^{-2}+2g_1^2)r^2} dr^2+\varphi_0^2 d\Omega_{3,k}^2+\Phi^2 dz^2
\cr && F=2\sqrt{k+g_1^2 \varphi_0^2}\,(\varphi_0^{-1}\Phi dt\wedge dr \wedge dz+\varphi_0^2 \Omega_{(3),k})
\eea
The (\ref{d6selfdual}) theory is a consistent truncation of the bosonic sector of $D=6$, ${\cal N}=(1,1)$ gauged supergravity that can be embedded in massive type IIA theory \cite{Cvetic:1999un}.  The effective cosmological constant in JT gravity in this embedding is related to the D4/D8-brane charges.

    Finally, it is worth noting that for $F=0$, the theory (\ref{theemd}) in general $D$ dimensions can
all be obtained from the circle reduction of Einstein gravity with a cosmological constant in $D+1$ dimensions \cite{Liu:2012ed}.

\section{Conclusions}
\label{sec:conclude}

In this paper, we demonstrated that JT gravity in two dimensions could be obtained from the consistent Kaluza-Klein reduction on a class of EMD theories (\ref{theemd}) in general $D$ dimensions.  For $D=4$ and 5, the EMD theories are truncations of the bosonic sector of supergravities.  This allows one to embed JT gravities in strings or M-theory, providing stringy interpretations of the SYK model.

The exact embeddings of this paper also allow one to understand the solutions of JT gravity in the light of higher dimensional theories.  For $g_1=0$, we find that a class of JT gravity solutions are related to the time-dependent extremal charged black holes in the EMD theories.  The solutions can be further lifted to become, for example, intersecting D1/D5-brane, where the worldsheet is the 2d-Milne universe instead of the more traditional 2d-Minkowski spacetime. They can also be lifted to the time-dependent M2/M2/M2 intersections. For $g_1\ne 0$, we find that the cosmological constant in JT gravity is related to D3-brane or D4/D8-brane charges depending on the specific route of the embeddings.

The exact embeddings of this paper imply that we can obtain the Schwarzian action directly in higher dimensions.  The subtlety is that the worldsheet or worldvolume of branes should be described by the Milner metric rather than the usual Minkowski metric. The fact that the Milne universe appears in the worldsheet or worldvolume is tantalizing.  Non-dilatonic extremal $p$-branes such as the M-branes or the D3-brane have in general  AdS$_d\times S^{D-d}$ as their near-horizon geometry.  When the boundary (the brane worldvolume) of the AdS$_d$ is the Milne universe, we have
\be
ds_d^2 = \fft{dr^2}{r^2} + r^2 (-dt^2 + t^2 d\Omega_{d-2,-1}^2) =\Big(
-r^2 dt^2 + \fft{dr^2}{r^2}\Big) + \Phi^2 d\Omega_{d-2,-1}^2\,,
\ee
where $d\Omega_{d-2,-1}^2$ is taken to be some compact metrics with negative cosmological constant.  The Kaluza-Klein reduction on $d\Omega_{d-2,-1}^2$ yields naturally a two-dimensional gravity with nearly AdS$_2$ vacuum geometry.  It would be of great interest to investigate the corresponding holographic NCFT$_1$.

\section*{Acknolwedgement}

Y.Z.L.~and H.L.~are supported in part by NSFC grants No.~11475024 and No.~11235003; S.L.L.~is supported in part by NSFC under Grants No.~11575022 and No.~11175016. We are grateful to Eoin Colg\'ain, Yong-Hui Qi, Zhao-Long Wang and Hossein Yavartanoo for useful discussions.

\appendix

\section{From JT gravity to SYK}
\label{jttosyk}

In this appendix, we give a detail review of how the SYK model can arise from JT gravity.  This review is based on the work of \cite{Maldacena:2016upp}. JT gravity admits the nearly AdS$_2$ solution (\ref{solution}).
In the Euclidean signature, the solution reads
\be
ds^2=\fft{dt^2+ dz^2}{z^2}\,,\qquad\Phi=\alpha z+\fft{\beta+\gamma t+\Lambda_{0}^2\alpha t^2}{z}
\label{solu}
\ee
where $z=1/r$ and we have chosen $\Lambda_0=-1$ without loss of generality.

For the theory to be well-defined, we need to include the Gibbons-Hawking surface term and the appropriate counterterm to the action(\ref{JT}). The total action
is given by (after setting $\Lambda_{0}=-1$):
\be
S=\fft{1}{16\pi}\int_{M} d^2x\sqrt{g}\Phi(R+2)+\fft{1}{8\pi}\int_{\partial M} \sqrt{h}\Phi (K-1)\,,
\ee
where $h$ is the induced metric on the relevant boundary.  Note that we also included the holographic counterterm above, as was done in \cite{Harlow:2018tqv}.

Following \cite{Maldacena:2016upp,Forste:2017kwy}, we slice the original asymptotic boundary in terms of a new parameter $u$. In other words, the new boundary deviates from the AdS fixed point and is parameterized by $(t(u),z(u))$. The variables $(t(u),z(u))$ are then interpreted as the dynamical fields of the boundary, where $u$ naturally serves as the Euclidean time of the boundary CFT. It is clear that the new boundary is fixed by
\be
g_{uu}=\dfrac{1}{\epsilon^2}=\dfrac{t'^2+z'^2}{z^2}\,,
\label{guu}
\ee
where a prime denotes a derivative with respect to $u$, and $\epsilon$ is an infinitesimal constant representing the UV cutoff. Eq.~(\ref{guu}) implies that $t(u)$ and $z(u)$ are not independent dynamic fields asymptotically, they are related by
\be
z(u)=\epsilon t'+\fft{1}{2}\epsilon^3~\fft{t''^2}{t'}+\cdots\,,\qquad \Phi\rightarrow\dfrac{\Phi_{{\rm inf}}}{\epsilon}\,.
\label{z}
\ee
where $\Phi_{{\rm inf}}$ is given by
\be
\Phi_{{\rm inf}}=\fft{\beta+\gamma t+\alpha t^2}{t'}\,.\label{Phiinf}
\ee
The rigid AdS geometry causes the bulk action to vanish identically, and the total action will be given by the boundary action, namely
\be
S=\fft{1}{8\pi}\int du\dfrac{\Phi_{{\rm inf}}}{\epsilon^2}(K-1)\,,\qquad K=h^{\mu\nu}\nabla_{\mu}n_{\nu}\,.
\label{efac}
\ee
To evaluate the vector $n=n^\mu \partial_\mu$ normal to the boundary, we note that the tangent vector is given by
\be
T=\fft{\partial}{\partial u}=t'\fft{\partial}{\partial t}+\epsilon t''\Big(1+\epsilon^2 (\fft{t'''}{t'}-\fft{1}{2}(\fft{t''}{t'})^2)\Big)\fft{\partial}{\partial z}\equiv T^\mu \partial_\mu\,,
\label{Tangent}
\ee
where we have used (\ref{z}). Since the normal vector is defined by
\be
T^{\mu}n_{\mu}=0\,,\qquad n_{\mu}n_{\nu}g^{\mu\nu}=1\,,
\ee
We find, up to the $\epsilon^2$ order, that
\be
n_{0}=\fft{t''}{t'^2}\big(1+\epsilon^2 {\rm Sch}(t(u),u)\big)\,,\qquad n_{1}=-\fft{1}{\epsilon t}\big(1-\epsilon^2 (\fft{t''}{t'})^2\big)\,,
\label{Normal}
\ee
where ${\rm Sch}(t(u),u)$ is the Schwarzian derivative
\be
{\rm Sch}(t(u),u)=\frac{2 t^{(3)} t'-3 t''^2}{2 t'^2}\,.
\ee
The normal vector $n^{\mu}$, the tangent vector $T^{\mu}$ and the metric $g_{\mu\nu}$ satisfy the identity
\be
g_{\mu\nu}-n_{\mu}n_{\nu}=\fft{T_{\mu}T_{\nu}}{T^2}\,.
\ee
It follows that the extrinsic curvature can be rewritten as
\be
K=\fft{T^{\mu}}{T^2}(n'_{\mu}-n_\rho\Gamma^{\rho}_{\mu\nu}T^{\nu})
\label{K}
\ee
Substituting (\ref{Tangent}) and (\ref{Normal}) into (\ref{K}), we find
\be
K=1+{\rm Sch}(t(u),u)\epsilon^2+\cdots\,,
\ee
where the dots represent those with higher powers of $\epsilon$.  Therefore, it is straightforward to see that  (\ref{efac}) is given by
\be
S=\fft{1}{8\pi}\int du~\Phi_{{\rm inf}}\,{\rm Sch}(t(u),u)\,.
\label{Schp}
\ee
From (\ref{Sch}), varying $t(u)$, we can obtain solution for $\Phi_{{\rm inf}}(u)$, i.e. (\ref{Phiinf}). It should be emphasized that $\Phi_{\rm inf}$ here should be viewed as a background rather than a dynamical field; it does not involve in the variation principle.  The equation (\ref{Phiinf}) is thus an equation of motion for $t(u)$ at a given background $\Phi_{\rm inf}(u)$.
Making a coordinate transformation
\be
\Phi_{{\rm inf}}\fft{\partial\tilde{u}}{\partial u}=\Phi_c\,,
\ee
where $\Phi_c$ is a constant, we find
\be
S=\fft{\Phi_c}{8\pi}\int d\tilde{u}\Big({\rm Sch}(t(\tilde{u}),\tilde{u})-{\rm Sch}(u(\tilde{u}),\tilde{u})\Big)\,.\label{schsch}
\ee
The second term in the bracket is a quantity that has no dynamical field and can be dropped, leading to the Schwarzian action
\be
S=\fft{\Phi_c}{8\pi}\int du\,{\rm Sch}(t(u),u)\,.
\label{Sch}
\ee
It is the effective action describing the IR behavior of the SYK model \cite{Maldacena:2016hyu}, and is invariant under the $SL(2,\mathbb R)$ transformation
\be
t(u)\rightarrow\dfrac{a\, t(u)+b}{c\, t(u)+d}\,,\qquad ad-bc=1\,.
\ee
It is of interest to note that had one considered $\Phi_{\rm inf}$ as a dynamical field, then eq.~(\ref{schsch}) would imply that the system had a ghost excitation.

\end{document}